\documentstyle[12pt,aps,prb]{revtex}

\def\PsfigVersion{1.9}
\ifx\undefined\psfig\else \fi

\let\LaTeXAtSign=\@
\let\@=\relax
\edef\psfigRestoreAt{\catcode`\@=\number\catcode`@\relax}
\catcode`\@=11\relax
\newwrite\@unused
\def\ps@typeout#1{{\let\protect\string\immediate\write\@unused{#1}}}
\ps@typeout{psfig/tex \PsfigVersion}

\def\figurepath{./}

\def\@nnil{\@nil}
\def\@empty{}
\def\@psdonoop#1\@@#2#3{}
\def\@psdo#1:=#2\do#3{\edef\@psdotmp{#2}\ifx\@psdotmp\@empty \else
    \expandafter\@psdoloop#2,\@nil,\@nil\@@#1{#3}\fi}
\def\@psdoloop#1,#2,#3\@@#4#5{\def#4{#1}\ifx #4\@nnil \else
       #5\def#4{#2}\ifx #4\@nnil \else#5\@ipsdoloop #3\@@#4{#5}\fi\fi}
\def\@ipsdoloop#1,#2\@@#3#4{\def#3{#1}\ifx #3\@nnil 
       \let\@nextwhile=\@psdonoop \else
      #4\relax\let\@nextwhile=\@ipsdoloop\fi\@nextwhile#2\@@#3{#4}}
\def\@tpsdo#1:=#2\do#3{\xdef\@psdotmp{#2}\ifx\@psdotmp\@empty \else
    \@tpsdoloop#2\@nil\@nil\@@#1{#3}\fi}
\def\@tpsdoloop#1#2\@@#3#4{\def#3{#1}\ifx #3\@nnil 
       \let\@nextwhile=\@psdonoop \else
      #4\relax\let\@nextwhile=\@tpsdoloop\fi\@nextwhile#2\@@#3{#4}}
\ifx\undefined\fbox
\newdimen\fboxrule
\newdimen\fboxsep
\newdimen\ps@tempdima
\newbox\ps@tempboxa
\long\def\fbox#1{\leavevmode\setbox\ps@tempboxa\hbox{#1}\ps@tempdima\fboxrule
    \advance\ps@tempdima \fboxsep \advance\ps@tempdima \dp\ps@tempboxa
   \hbox{\lower \ps@tempdima\hbox
  {\vbox{\hrule height \fboxrule
          \hbox{\vrule width \fboxrule \hskip\fboxsep
          \vbox{\vskip\fboxsep \box\ps@tempboxa\vskip\fboxsep}\hskip 
                 \fboxsep\vrule width \fboxrule}
                 \hrule height \fboxrule}}}}
\fi
\newread\ps@stream
\newif\ifnot@eof       
\newif\if@noisy        
\newif\if@atend        
\newif\if@psfile       
{\catcode`\%=12\global\gdef\epsf@start{
\def\epsf@PS{PS}
\def\epsf@getbb#1{%
\openin\ps@stream=#1
\ifeof\ps@stream\ps@typeout{Error, File #1 not found}\else
   {\not@eoftrue \chardef\other=12
    \def\do##1{\catcode`##1=\other}\dospecials \catcode`\ =10
    \loop
       \if@psfile
	  \read\ps@stream to \epsf@fileline
       \else{
	  \obeyspaces
          \read\ps@stream to \epsf@tmp\global\let\epsf@fileline\epsf@tmp}
       \fi
       \ifeof\ps@stream\not@eoffalse\else
       \if@psfile\else
       \expandafter\epsf@test\epsf@fileline:. \\%
       \fi
          \expandafter\epsf@aux\epsf@fileline:. \\%
       \fi
   \ifnot@eof\repeat
   }\closein\ps@stream\fi}%
\long\def\epsf@test#1#2#3:#4\\{\def\epsf@testit{#1#2}
			\ifx\epsf@testit\epsf@start\else
\ps@typeout{Warning! File does not start with `\epsf@start'.  It may not be a PostScript file.}
			\fi
			\@psfiletrue} 
{\catcode`\%=12\global\let\epsf@percent=
\long\def\epsf@aux#1#2:#3\\{\ifx#1\epsf@percent
   \def\epsf@testit{#2}\ifx\epsf@testit\epsf@bblit
	\@atendfalse
        \epsf@atend #3 . \\%
	\if@atend	
	   \if@verbose{
		\ps@typeout{psfig: found `(atend)'; continuing search}
	   }\fi
        \else
        \epsf@grab #3 . . . \\%
        \not@eoffalse
        \global\no@bbfalse
        \fi
   \fi\fi}%
\def\epsf@grab #1 #2 #3 #4 #5\\{%
   \global\def\epsf@llx{#1}\ifx\epsf@llx\empty
      \epsf@grab #2 #3 #4 #5 .\\\else
   \global\def\epsf@lly{#2}%
   \global\def\epsf@urx{#3}\global\def\epsf@ury{#4}\fi}%
\def\epsf@atendlit{(atend)} 
\def\epsf@atend #1 #2 #3\\{%
   \def\epsf@tmp{#1}\ifx\epsf@tmp\empty
      \epsf@atend #2 #3 .\\\else
   \ifx\epsf@tmp\epsf@atendlit\@atendtrue\fi\fi}

\chardef\psletter = 11 
\chardef\other = 12

\newif \ifdebug 
\newif\ifc@mpute 
\c@mputetrue 

\let\then = \relax
\def\r@dian{pt }
\let\r@dians = \r@dian
\let\dimensionless@nit = \r@dian
\let\dimensionless@nits = \dimensionless@nit
\def\internal@nit{sp }
\let\internal@nits = \internal@nit
\newif\ifstillc@nverging
\def \Mess@ge #1{\ifdebug \then \message {#1} \fi}

{ 
	\catcode `\@ = \psletter
	\gdef \nodimen {\expandafter \n@dimen \the \dimen}
	\gdef \term #1 #2 #3%
	       {\edef \t@ {\the #1}
		\edef \t@@ {\expandafter \n@dimen \the #2\r@dian}%
		\t@rm {\t@} {\t@@} {#3}%
	       }
	\gdef \t@rm #1 #2 #3%
	       {{%
		\count 0 = 0
		\dimen 0 = 1 \dimensionless@nit
		\dimen 2 = #2\relax
		\Mess@ge {Calculating term #1 of \nodimen 2}%
		\loop
		\ifnum	\count 0 < #1
		\then	\advance \count 0 by 1
			\Mess@ge {Iteration \the \count 0 \space}%
			\Multiply \dimen 0 by {\dimen 2}%
			\Mess@ge {After multiplication, term = \nodimen 0}%
			\Divide \dimen 0 by {\count 0}%
			\Mess@ge {After division, term = \nodimen 0}%
		\repeat
		\Mess@ge {Final value for term #1 of 
				\nodimen 2 \space is \nodimen 0}%
		\xdef \Term {#3 = \nodimen 0 \r@dians}%
		\aftergroup \Term
	       }}
	\catcode `\p = \other
	\catcode `\t = \other
	\gdef \n@dimen #1pt{#1} 
}

\def \Divide #1by #2{\divide #1 by #2} 

\def \Multiply #1by #2
       {{
	\count 0 = #1\relax
	\count 2 = #2\relax
	\count 4 = 65536
	\Mess@ge {Before scaling, count 0 = \the \count 0 \space and
			count 2 = \the \count 2}%
	\ifnum	\count 0 > 32767 
	\then	\divide \count 0 by 4
		\divide \count 4 by 4
	\else	\ifnum	\count 0 < -32767
		\then	\divide \count 0 by 4
			\divide \count 4 by 4
		\else
		\fi
	\fi
	\ifnum	\count 2 > 32767 
	\then	\divide \count 2 by 4
		\divide \count 4 by 4
	\else	\ifnum	\count 2 < -32767
		\then	\divide \count 2 by 4
			\divide \count 4 by 4
		\else
		\fi
	\fi
	\multiply \count 0 by \count 2
	\divide \count 0 by \count 4
	\xdef \product {#1 = \the \count 0 \internal@nits}%
	\aftergroup \product
       }}

\def\r@duce{\ifdim\dimen0 > 90\r@dian \then   
		\multiply\dimen0 by -1
		\advance\dimen0 by 180\r@dian
		\r@duce
	    \else \ifdim\dimen0 < -90\r@dian \then  
		\advance\dimen0 by 360\r@dian
		\r@duce
		\fi
	    \fi}

\def\Sine#1%
       {{%
	\dimen 0 = #1 \r@dian
	\r@duce
	\ifdim\dimen0 = -90\r@dian \then
	   \dimen4 = -1\r@dian
	   \c@mputefalse
	\fi
	\ifdim\dimen0 = 90\r@dian \then
	   \dimen4 = 1\r@dian
	   \c@mputefalse
	\fi
	\ifdim\dimen0 = 0\r@dian \then
	   \dimen4 = 0\r@dian
	   \c@mputefalse
	\fi
	\ifc@mpute \then
		\divide\dimen0 by 180
		\dimen0=3.141592654\dimen0
		\dimen 2 = 3.1415926535897963\r@dian 
		\divide\dimen 2 by 2 
		\Mess@ge {Sin: calculating Sin of \nodimen 0}%
		\count 0 = 1 
		\dimen 2 = 1 \r@dian 
		\dimen 4 = 0 \r@dian 
		\loop
			\ifnum	\dimen 2 = 0 
			\then	\stillc@nvergingfalse 
			\else	\stillc@nvergingtrue
			\fi
			\ifstillc@nverging 
			\then	\term {\count 0} {\dimen 0} {\dimen 2}%
				\advance \count 0 by 2
				\count 2 = \count 0
				\divide \count 2 by 2
				\ifodd	\count 2 
				\then	\advance \dimen 4 by \dimen 2
				\else	\advance \dimen 4 by -\dimen 2
				\fi
		\repeat
	\fi		
			\xdef \sine {\nodimen 4}%
       }}

\def\Cosine#1{\ifx\sine\UnDefined\edef\Savesine{\relax}\else
		             \edef\Savesine{\sine}\fi
	{\dimen0=#1\r@dian\advance\dimen0 by 90\r@dian
	 \Sine{\nodimen 0}
	 \xdef\cosine{\sine}
	 \xdef\sine{\Savesine}}}	      

\def\psdraft{
	\def\@psdraft{0}
}
\def\psfull{
	\def\@psdraft{100}
}

\psfull

\newif\if@scalefirst
\def\psscalefirst{\@scalefirsttrue}
\def\psrotatefirst{\@scalefirstfalse}
\psrotatefirst

\newif\if@draftbox
\def\psnodraftbox{
	\@draftboxfalse
}
\def\psdraftbox{
	\@draftboxtrue
}
\@draftboxtrue

\newif\if@prologfile
\newif\if@postlogfile
\def\pssilent{
	\@noisyfalse
}
\def\psnoisy{
	\@noisytrue
}
\psnoisy
\newif\if@bbllx
\newif\if@bblly
\newif\if@bburx
\newif\if@bbury
\newif\if@height
\newif\if@width
\newif\if@rheight
\newif\if@rwidth
\newif\if@angle
\newif\if@clip
\newif\if@verbose
\newif\if@scale
\def\@p@@sclip#1{\@cliptrue}

\newif\if@decmpr

\def\@p@@sfigure#1{\def\@p@sfile{null}\def\@p@sbbfile{null}
	        \openin1=#1.bb
		\ifeof1\closein1
	        	\openin1=\figurepath#1.bb
			\ifeof1\closein1
			        \openin1=#1
				\ifeof1\closein1%
				       \openin1=\figurepath#1
					\ifeof1
					   \ps@typeout{Error, File #1 not found}
						\if@bbllx\if@bblly
				   		\if@bburx\if@bbury
			      				\def\@p@sfile{#1}%
			      				\def\@p@sbbfile{#1}%
							\@decmprfalse
				  	   	\fi\fi\fi\fi
					\else\closein1
				    		\def\@p@sfile{\figurepath#1}%
				    		\def\@p@sbbfile{\figurepath#1}%
						\@decmprfalse
	                       		\fi%
			 	\else\closein1%
					\def\@p@sfile{#1}
					\def\@p@sbbfile{#1}
					\@decmprfalse
			 	\fi
			\else
				\def\@p@sfile{\figurepath#1}
				\def\@p@sbbfile{\figurepath#1.bb}
				\@decmprtrue
			\fi
		\else
			\def\@p@sfile{#1}
			\def\@p@sbbfile{#1.bb}
			\@decmprtrue
		\fi}

\def\@p@@sfile#1{\@p@@sfigure{#1}}

\def\@p@@sbbllx#1{
		\@bbllxtrue
		\dimen100=#1
		\edef\@p@sbbllx{\number\dimen100}
}
\def\@p@@sbblly#1{
		\@bbllytrue
		\dimen100=#1
		\edef\@p@sbblly{\number\dimen100}
}
\def\@p@@sbburx#1{
		\@bburxtrue
		\dimen100=#1
		\edef\@p@sbburx{\number\dimen100}
}
\def\@p@@sbbury#1{
		\@bburytrue
		\dimen100=#1
		\edef\@p@sbbury{\number\dimen100}
}
\def\@p@@sheight#1{
		\@heighttrue
		\dimen100=#1
   		\edef\@p@sheight{\number\dimen100}
}
\def\@p@@swidth#1{
		\@widthtrue
		\dimen100=#1
		\edef\@p@swidth{\number\dimen100}
}
\def\@p@@srheight#1{
		\@rheighttrue
		\dimen100=#1
		\edef\@p@srheight{\number\dimen100}
}
\def\@p@@srwidth#1{
		\@rwidthtrue
		\dimen100=#1
		\edef\@p@srwidth{\number\dimen100}
}
\def\@p@@sangle#1{
		\@angletrue
		\edef\@p@sangle{#1} 
}
\def\@p@@srotate#1{\@p@@sangle{-#1}}
\def\@p@@sscale#1{
		\@scaletrue
		\edef\@p@sscale{#1}
}
\def\@p@@ssilent#1{ 
		\@verbosefalse
}
\def\@p@@sprolog#1{\@prologfiletrue\def\@prologfileval{#1}}
\def\@p@@spostlog#1{\@postlogfiletrue\def\@postlogfileval{#1}}
\def\@cs@name#1{\csname #1\endcsname}
\def\@setparms#1=#2,{\@cs@name{@p@@s#1}{#2}}
\def\ps@init@parms{
		\@bbllxfalse \@bbllyfalse
		\@bburxfalse \@bburyfalse
		\@heightfalse \@widthfalse
		\@rheightfalse \@rwidthfalse
		\@scalefalse
		\def\@p@sbbllx{}\def\@p@sbblly{}
		\def\@p@sbburx{}\def\@p@sbbury{}
		\def\@p@sheight{}\def\@p@swidth{}
		\def\@p@srheight{}\def\@p@srwidth{}
		\def\@p@sangle{0}
		\def\@p@sfile{} \def\@p@sbbfile{}
		\def\@p@scost{10}
		\def\@sc{}
		\@prologfilefalse
		\@postlogfilefalse
		\@clipfalse
		\if@noisy
			\@verbosetrue
		\else
			\@verbosefalse
		\fi
}
\def\parse@ps@parms#1{
	 	\@psdo\@psfiga:=#1\do
		   {\expandafter\@setparms\@psfiga,}}
\newif\ifno@bb
\def\bb@missing{
	\if@verbose{
		\ps@typeout{psfig: searching \@p@sbbfile \space  for bounding box}
	}\fi
	\no@bbtrue
	\epsf@getbb{\@p@sbbfile}
        \ifno@bb \else \bb@cull\epsf@llx\epsf@lly\epsf@urx\epsf@ury\fi
}	
\def\bb@cull#1#2#3#4{
	\dimen100=#1 bp\edef\@p@sbbllx{\number\dimen100}
	\dimen100=#2 bp\edef\@p@sbblly{\number\dimen100}
	\dimen100=#3 bp\edef\@p@sbburx{\number\dimen100}
	\dimen100=#4 bp\edef\@p@sbbury{\number\dimen100}
	\no@bbfalse
}
\newdimen\p@intvaluex
\newdimen\p@intvaluey
\def\rotate@#1#2{{\dimen0=#1 sp\dimen1=#2 sp
		  \global\p@intvaluex=\cosine\dimen0
		  \dimen3=\sine\dimen1
		  \global\advance\p@intvaluex by -\dimen3
		  \global\p@intvaluey=\sine\dimen0
		  \dimen3=\cosine\dimen1
		  \global\advance\p@intvaluey by \dimen3
		  }}
\def\compute@bb{
		\no@bbfalse
		\if@bbllx \else \no@bbtrue \fi
		\if@bblly \else \no@bbtrue \fi
		\if@bburx \else \no@bbtrue \fi
		\if@bbury \else \no@bbtrue \fi
		\ifno@bb \bb@missing \fi
		\ifno@bb \ps@typeout{FATAL ERROR: no bb supplied or found}
			\no-bb-error
		\fi
		\count203=\@p@sbburx
		\count204=\@p@sbbury
		\advance\count203 by -\@p@sbbllx
		\advance\count204 by -\@p@sbblly
		\edef\ps@bbw{\number\count203}
		\edef\ps@bbh{\number\count204}
		\if@angle 
			\Sine{\@p@sangle}\Cosine{\@p@sangle}
	        	{\dimen100=\maxdimen\xdef\r@p@sbbllx{\number\dimen100}
					    \xdef\r@p@sbblly{\number\dimen100}
			                    \xdef\r@p@sbburx{-\number\dimen100}
					    \xdef\r@p@sbbury{-\number\dimen100}}
                        \def\minmaxtest{
			   \ifnum\number\p@intvaluex<\r@p@sbbllx
			      \xdef\r@p@sbbllx{\number\p@intvaluex}\fi
			   \ifnum\number\p@intvaluex>\r@p@sbburx
			      \xdef\r@p@sbburx{\number\p@intvaluex}\fi
			   \ifnum\number\p@intvaluey<\r@p@sbblly
			      \xdef\r@p@sbblly{\number\p@intvaluey}\fi
			   \ifnum\number\p@intvaluey>\r@p@sbbury
			      \xdef\r@p@sbbury{\number\p@intvaluey}\fi
			   }
			\rotate@{\@p@sbbllx}{\@p@sbblly}
			\minmaxtest
			\rotate@{\@p@sbbllx}{\@p@sbbury}
			\minmaxtest
			\rotate@{\@p@sbburx}{\@p@sbblly}
			\minmaxtest
			\rotate@{\@p@sbburx}{\@p@sbbury}
			\minmaxtest
			\edef\@p@sbbllx{\r@p@sbbllx}\edef\@p@sbblly{\r@p@sbblly}
			\edef\@p@sbburx{\r@p@sbburx}\edef\@p@sbbury{\r@p@sbbury}
		\fi
		\count203=\@p@sbburx
		\count204=\@p@sbbury
		\advance\count203 by -\@p@sbbllx
		\advance\count204 by -\@p@sbblly
		\edef\@bbw{\number\count203}
		\edef\@bbh{\number\count204}
}
\def\in@hundreds#1#2#3{\count240=#2 \count241=#3
		     \count100=\count240	
		     \divide\count100 by \count241
		     \count101=\count100
		     \multiply\count101 by \count241
		     \advance\count240 by -\count101
		     \multiply\count240 by 10
		     \count101=\count240	
		     \divide\count101 by \count241
		     \count102=\count101
		     \multiply\count102 by \count241
		     \advance\count240 by -\count102
		     \multiply\count240 by 10
		     \count102=\count240	
		     \divide\count102 by \count241
		     \count200=#1\count205=0
		     \count201=\count200
			\multiply\count201 by \count100
		 	\advance\count205 by \count201
		     \count201=\count200
			\divide\count201 by 10
			\multiply\count201 by \count101
			\advance\count205 by \count201
		     \count201=\count200
			\divide\count201 by 100
			\multiply\count201 by \count102
			\advance\count205 by \count201
		     \edef\@result{\number\count205}
}
\def\ps@scaleinhundreds#1{
		\in@hundreds{#1}{\@p@sscale}{100}
		\edef#1{\@result}
}
\def\compute@wfromh{
		\in@hundreds{\@p@sheight}{\@bbw}{\@bbh}
		\edef\@p@swidth{\@result}
}
\def\compute@hfromw{
	        \in@hundreds{\@p@swidth}{\@bbh}{\@bbw}
		\edef\@p@sheight{\@result}
}
\def\compute@handw{
		\if@height 
			\if@width
			\else
				\compute@wfromh
			\fi
		\else 
			\if@width
				\compute@hfromw
			\else
				\edef\@p@sheight{\@bbh}
				\edef\@p@swidth{\@bbw}
			\fi
		\fi
}
\def\compute@resv{
		\if@rheight \else \edef\@p@srheight{\@p@sheight} \fi
		\if@rwidth \else \edef\@p@srwidth{\@p@swidth} \fi
}
\def\compute@sizes{
	\compute@bb
	\if@scalefirst\if@angle
	\if@width
	   \in@hundreds{\@p@swidth}{\@bbw}{\ps@bbw}
	   \edef\@p@swidth{\@result}
	\fi
	\if@height
	   \in@hundreds{\@p@sheight}{\@bbh}{\ps@bbh}
	   \edef\@p@sheight{\@result}
	\fi
	\fi\fi
	\compute@handw
	\compute@resv
	\if@scale
	   \if@verbose
	      \ps@typeout{(scaling by \@p@sscale)}%
	   \fi
	   \ps@scaleinhundreds{\@p@swidth}%
	   \ps@scaleinhundreds{\@p@sheight}%
	   \ps@scaleinhundreds{\@p@srwidth}%
	   \ps@scaleinhundreds{\@p@srheight}%
	\fi
}

\def\psfig#1{\vbox {
	\ps@init@parms
	\parse@ps@parms{#1}
	\compute@sizes
	\ifnum\@p@scost<\@psdraft{
		\special{ps::[begin] 	\@p@swidth \space \@p@sheight \space
				\@p@sbbllx \space \@p@sbblly \space
				\@p@sbburx \space \@p@sbbury \space
				startTexFig \space }
		\if@angle
			\special {ps:: \@p@sangle \space rotate \space} 
		\fi
		\if@clip{
			\if@verbose{
				\ps@typeout{(clip)}
			}\fi
			\special{ps:: doclip \space }
		}\fi
		\if@prologfile
		    \special{ps: plotfile \@prologfileval \space } \fi
		\if@decmpr{
			\if@verbose{
				\ps@typeout{psfig: including \@p@sfile.Z \space }
			}\fi
			\special{ps: plotfile "`zcat \@p@sfile.Z" \space }
		}\else{
			\if@verbose{
				\ps@typeout{psfig: including \@p@sfile \space }
			}\fi
			\special{ps: plotfile \@p@sfile \space }
		}\fi
		\if@postlogfile
		    \special{ps: plotfile \@postlogfileval \space } \fi
		\special{ps::[end] endTexFig \space }
		\vbox to \@p@srheight true sp{
			\hbox to \@p@srwidth true sp{
				\hss
			}
		\vss
		}
	}\else{
		\if@draftbox{		
			\hbox{\frame{\vbox to \@p@srheight true sp{
			\vss
			\hbox to \@p@srwidth true sp{ \hss \@p@sfile \hss }
			\vss
			}}}
		}\else{
			\vbox to \@p@srheight true sp{
			\vss
			\hbox to \@p@srwidth true sp{\hss}
			\vss
			}
		}\fi

	}\fi
}}
\psfigRestoreAt
\let\@=\LaTeXAtSign

\def\al2o3{Al$_2$O$_3$}
\def\aal2o3{$\alpha$-Al$_2$O$_3$}

\begin{document}
\draft
\tightenlines

\title{The prismatic $\Sigma 3$ $(10\bar{1}0)$ twin boundary in \aal2o3
       investigated by density functional theory and transmission electron
       microscopy}
\author{Stefano Fabris, Stefan Nufer,\footnote{Present address: Robert
        Bosch GmbH, Robert-Bosch-Platz 1, D-70839 Gerlingen-Schillerh\"ohe,
        Germany.} and Christian Els\"asser} 
\address{Max-Planck-Institut f\"ur Metallforschung, Seestrasse 92,
         D-70174 Stuttgart, Germany}
\date{\today} 
\maketitle

\begin{center}
\begin{abstract}

The microscopic structure of the prismatic $\Sigma 3$ $(10\bar{1}0)$ twin
boundary in \aal2o3 is characterized theoretically by ab-initio
local-density-functional theory, and experimentally by spatial-resolution
electron energy-loss spectroscopy in a scanning transmission electron
microscope (STEM), measuring energy-loss near-edge structures (ELNES) of
the oxygen $K$-ionization edge.  Theoretically, two distinct microscopic
variants for this twin interface with low interface energies are derived
and analysed.  Experimentally, it is demonstrated that the spatial and
energetical resolutions of present high-performance STEM instruments are
insufficient to discriminate the subtle differences of the two proposed
interface variants. It is predicted that for the currently developed next
generation of analytical electron microscopes the prismatic twin
interface will provide a promising benchmark case to demonstrate the
achievement of ELNES with spatial resolution of individual atom columns.

\end{abstract}

\end{center}


\section{Introduction}

  The local atomistic and electronic structures at extended defects in
  materials, such as grain boundaries in polycrystals or interfaces in
  layered nanostructures, are highly influential on the structural or
  functional properties of many technological devices, and hitherto, strong
  efforts have been made to characterize them microscopically. Recently,
  significant progress has been achieved by combining experimental imaging
  and spectroscopy with theoretical ab initio calculations. On the
  experimental side, imaging and electron-energy-loss spectroscopy (EELS)
  are combined via transmission electron microscopy (TEM) with sub-nm
  spatial and sub-eV energetical resolutions. On the theoretical side, the
  first-principles electronic-structure calculations are based on the local
  density functional theory (LDFT). This current progress of TEM and LDFT
  has been demonstrated for instance for the rhombohedral $\Sigma$7
  ($\bar{1}012$) twin boundary in \aal2o3,~\cite{Nufer01,Marinop00} a model
  case of a realistic interface in a technologically relevant ceramic
  material.


  In the case of the rhombohedral twin boundary,~\cite{Nufer01} both the
  quantitative analysis of high-resolution TEM (HRTEM) and EELS were able to
  discriminate between the lowest-energy interface model and another
  metastable one, which were predicted by ab-initio LDFT
  calculations.~\cite{Marinop00} On the one side, for the high-resolution
  TEM combined with image simulation, the discrimination was at the limits
  of the present experimental resolution and simulation accuracy. On the
  other side, the electron-energy-loss spectroscopy in a scanning
  transmission electron microscope (STEM) provided spatially resolved
  energy-loss near-edge structures (ELNES) in the oxygen $K$-ionization
  edge, which, in combination with calculated local densities of unoccupied
  one-electron states, yielded a more clear distinction between the
  lowest-energy and metastable higher-energy interface models.


  In the present work, this effort in microscopic interface analysis is
  extended to the prismatic $\Sigma$3 ($10\bar{1}0$) twin of \aal2o3.  In
  this case, the theoretical LDFT analysis results again in two metastable
  low-energy interface models. In distinction from the rhombohedral-twin,
  it is demonstrated that experimental EELS in a STEM with spatial
  resolution of 1$-$2~nm, as in Ref. \onlinecite{Nufer01}, is {\it not}
  sufficient to discriminate the ELNES of the theoretically predicted
  models for the prismatic twin interface. It is predicted that for a new
  generation of analytical TEM instruments with spatial resolution of
  0.1$-$0.2~nm, which are currently under development in academic and
  industrial TEM research laboratories,~\cite{Benthem02} the prismatic twin
  interface will provide a promising benchmark for a successful achievement
  of atom-column-resolution ELNES, via a site-specific comparison of
  measured oxygen $K$-ionization edges and calculated local densities of
  states.  A further challenge for quantitative high-resolution TEM image
  analysis is provided by the calculated interfacial arrangement of atoms.

  After a concise summary of the computational and experimental methods
  (Sec. II), we will report the search procedure for metastable
  variants of the twin boundary via empirical shell-model and ab-initio
  LDFT calculations (Sec. IIIA), the ab-initio LDFT analysis of the
  interfacial atomistics and energetics (Sec. IIIB), the calculated local
  densities of states and experimental ELNES (Sec. IIIC), and the simulated
  HRTEM images of the theoretically obtained two variants of the twin
  boundary (Sec. IIID). The main conclusions will be summarized in Sec. IV.


\section{Methods}

\subsection{Computational approaches}

 The theoretical analysis described in this work combines empirical
 atomistic modeling, based on an ionic shell-model potential,~\cite{Dick58}
 and first-principles electronic-structure calculations, based on the
 density-functional theory~\cite{Hohenberg64} in the local-density
 approximation for exchange and correlation~\cite{Kohn65} (local density
 functional theory, LDFT).

 The shell-model potential was parameterized for \aal2o3 by Lewis and
 Catlow~\cite{Lewis85} via fitting to experimental and theoretical
 data. The potential describes the crystal as fully ionic, with long-range
 electrostatic interactions (point charges on both atomic sites and charged
 shells on the anion sites), short-range overlap repulsion, and weak van
 der Waals interactions. The model reproduces quite well the structural
 properties of bulk and defective \aal2o3, but fails to predict the
 energetic stability of the corundum structure relative to other possible
 polymorphs.~\cite{Gale92,Wilson96al2o3} In this work, the model is applied
 for a qualitative search of metastable translation states of \aal2o3
 (corundum) bicrystals, hence it is applied where its predictive power is
 well justified. The atomistic interface-structure variations with the
 shell-model potential are performed with the computer code
 GULP.~\cite{Gale97}

 In our first-principles mixed-basis pseudopotential LDFT
 approach,~\cite{Elsaesser90,Ho92,ElsaesserTh,MeyerTh,MeyerUnp} the crystal
 valence wave functions are represented by a mixed basis set, consisting of
 plane waves and localized functions. In the specific case of \al2o3, the
 plane waves are limited by the maximum kinetic energy E$_{pw}$=300
 eV. Three localized functions with angular momentum $l=1$, $m=0,\pm 1$,
 are centered at each oxygen site, and spatially confined by a sphere of
 radius r$_{lo}$=1.16 \AA. The interactions of the atomic nuclei and core
 electrons on the valence electrons are described with norm-conserving
 non-local ionic pseudopotentials. More details on the construction of the
 pseudopotentials, and on the predictions of the method for the structural
 parameters of bulk \aal2o3 can be found in
 Ref.~\onlinecite{Marinop00}. The total energies for the prismatic twins
 are calculated by sampling the Brillouin zone of orthorhombic 120-atom
 supercells (see below) on a 2x1x1 {\bf k}-point mesh.  The forces on all
 atoms are explicitly calculated after self-consistency of the electronic
 structure is achieved, and thus the method allows for efficient structural
 relaxation of the atomic positions.

\begin{figure}
\caption{Corundum structure of \aal2o3 viewed from the [$ 1\bar{2}10$]
direction. Light-gray and black circles represent oxygen and aluminium
atoms, respectively. The dashed line marks the ($10\bar{1}0$) plane
leading to the prismatic $\Sigma 3$ interface.}
\label{cryst}
\centerline{\psfig{file=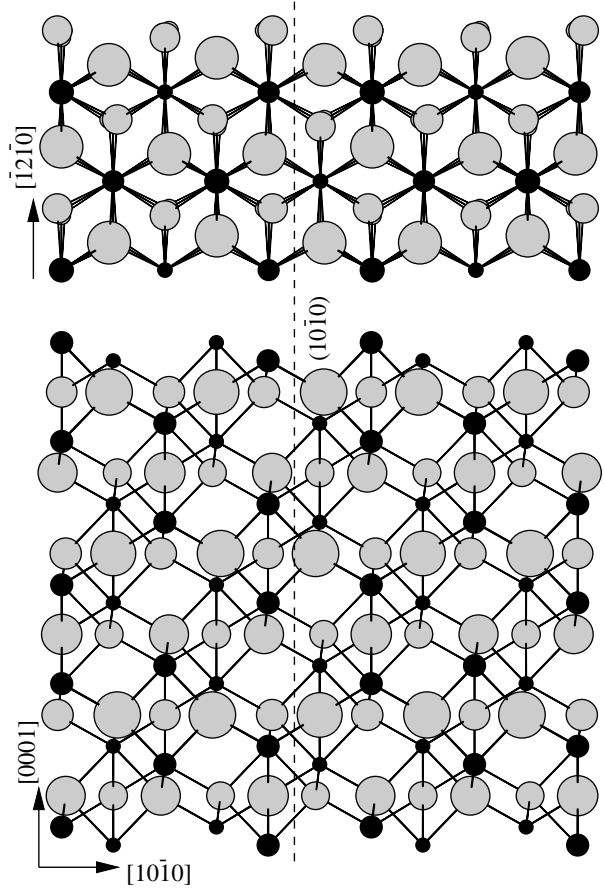,width=8cm,angle=0}} 
\end{figure}

 The crystal structure of \aal2o3 (corundum or sapphire, space group
 $R\bar{3}c$), projected on the ($\bar{1}2\bar{1}0$) plane, is shown in
 Fig.~\ref{cryst}: light and dark circles represent oxygen and aluminium
 ions, respectively, and the dashed line marks a prismatic ($10\bar{1}0$)
 plane. The twin boundaries are modeled with 120-atom orthorhombic
 supercells. These are defined by the lattice vectors {\bf
 e$_1$}=[$\bar{1}2\bar{1}0$], {\bf e$_2$}=[0001], and {\bf
 e$_3$}=[$10\bar{1}0$]. The supercell contains two grains of \aal2o3 with a
 misorientation given by ($10\bar{1}0$)$||$($10\bar{1}0$) and
 [$1\bar{2}10$]$||$[$\bar{1}2\bar{1}0$]: the three-dimensional periodic
 repetition generates a sequence of equivalent interfaces, separated by
 6.8 {\AA} thick slabs of bulk corundum.


\subsection{Experimental procedures}

 The prismatic twin bicrystal of \aal2o3 was prepared by W. Kurtz, MPI
 f\"{u}r Metallforschung Stuttgart, by diffusion
 bonding~\cite{Fischmeister93} of two halves of highly pure single crystals
 of sapphire in ultra-high vacuum. The resulting bicrystal has a deviation
 of $\approx$0.1$^\circ$ from the perfect misorientation (measured by
 electron and X-ray diffraction). The impurity concentration at the
 interface was below the detection limit of energy-dispersive X-ray (EDX)
 spectroscopy (0.3 atoms/nm$^2$).~\cite{NuferTh} The sample preparation for
 transmission electron microscopy and spectroscopy involved conventional
 sawing, polishing, grinding, and finally ion milling at rather low energy
 of 2.5 keV, which resulted in highly radiation-damage resistant TEM
 samples.~\cite{NuferTh} Electron energy-loss spectra (EELS) were recorded
 in a dedicated scanning transmission electron microscope (STEM, VG
 HB501UX) run at 100 kV and equipped with a parallel EELS spectrometer and
 an EDX detector. The measurements of spatially resolved energy-loss
 near-edge structures (ELNES) of the oxygen $K$ edge were performed in a
 spot modus, with energy dispersion of 0.1 eV per channel and effective
 electron spot diameter between 1 and 2 nm.  The energy resolution of the
 system was 0.8 eV. Extensive experimental TEM investigations of
 diffusion-bonded twin bicrystals of \aal2o3 with various misorientations
 are reported in Ref.~\onlinecite{NuferTh}.
 
\section{Presentation and Discussion of Results}

\subsection{Metastable variants of the prismatic twin interface}

 An initial search for metastable lateral translation states of the
 prismatic twin bicrystal was done by modeling the two grains as rigid
 blocks, and exploring the energy difference as a function of the relative
 rigid shift in the boundary plane ($10\bar{1}0$). A complete structural
 minimization was then performed only for the resulting equilibrium
 bicrystal configurations. The energetics was calculated via atomistic
 modeling based on an empirical shell-model potential.~\cite{Lewis85} This
 simple but effective method has hitherto always succeeded in isolating the
 complete set of possible metastable translation states of the alumina
 bicrystals,~\cite{Marinop00,Marinop01,sigma13} which is the only goal of
 this part of the study.

\begin{figure}
\caption{Atomic structure of one grain projected on the prismatic plane
($ 10\bar{1}0$). The in-plane extension of the computational cell is
marked by the lattice vectors ${\bf e_1} = [\bar{1}2\bar{1}0]$ and ${\bf
e_2} = [0001]$, and by solid lines. M, G and S denote the translation
states of the upper grain, leading respectively to Mirror, Glide-mirror
and Screw-rotation twins. The atoms are represented as in
Fig.~\ref{cryst}.}
\label{prism}
\centerline{\psfig{file=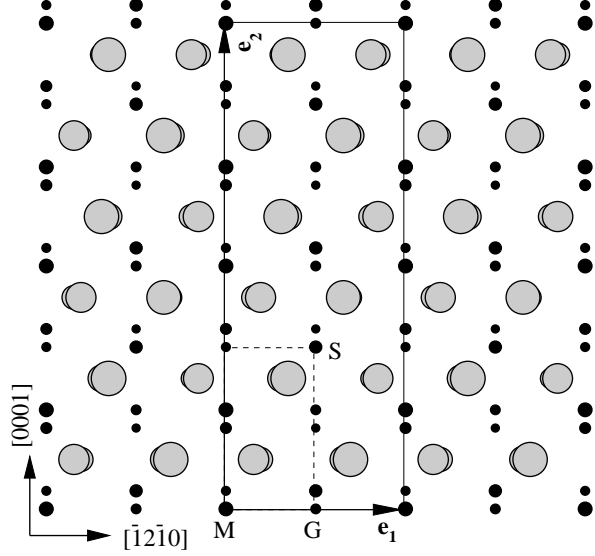,width=8cm,angle=0}} 
\end{figure}

 The lateral translation state $T=T_1 {\bf e_1}+ T_2 {\bf e_2}$ is defined
 in terms of the lattice vectors ${\bf e_1} = [\bar{1}2\bar{1}0]$ and ${\bf
 e_2} = [0001]$. The reference bicrystal configuration, corresponding to
 ($T_1,T_2$)=(0,0), was constructed from a block of bulk alumina by
 rotating one half of the crystal by 180$^\circ$ around $[10\bar{1}0]$ with
 respect to the other half. Since the rotation axis contains an inversion
 point of the crystal, the rotation generates a mirror-symmetric twin
 boundary, denoted by M hereafter. Fig.~\ref{prism} shows the
 projection of one grain on the prismatic ($10\bar{1}0$) plane: the solid
 lines correspond to the lattice vectors ${\bf e_1}$ and ${\bf e_2}$
 identifying the cell-edges of the orthorhombic supercell used in the
 calculations. The label M marks the position of the other grain in the
 reference M configuration, where the interfacial atoms of the two grains
 face each other at very close distance.

\begin{figure}
\caption{Shell-model total energy (Ry/supercell) for the twin interfaces
as a function of the lateral translation state $T_1{\bf e}_1+T_2{\bf
e}_2$, with {\bf e}$_1$=[$\bar{1}2\bar{1}0$] and {\bf e}$_2$=[$0001$].}
\label{shell}
 \centerline{\psfig{file=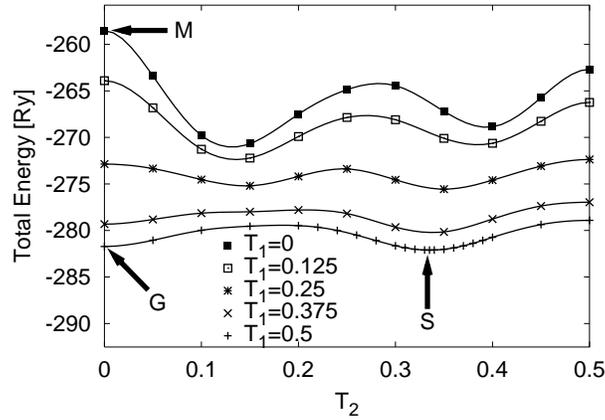,width=8cm,angle=-90}} 
\end{figure}

 The energetics of the bicrystal as a function of the translation state
 $(T_1,T_2)$ is shown in Fig.~\ref{shell}. Note how the highest energies
 correspond to the energy profile for $T_1$=0 (showing only two saddle
 points), while the lowest energies correspond to the energy profile for
 $T_1$=1/2. This can be easily understood by noting that the translation
 states with $T_1$=1/2 reconstruct a highly ordered bulk-like atomic
 environment with hexagonal symmetry in the [0001] projection. Only two
 metastable boundary variants are identified, with translation states
 $(T_1,T_2)$=(1/2,0) and $(T_1,T_2)$=(1/2,1/3), denoted by G and S,
 respectively. In the most stable S variant, the two grains are related by
 a screw-rotation symmetry operation around ${\bf e_3}$=[$10\bar{1}0$],
 while a glide-mirror symmetry operation with respect to ($10\bar{1}0$)
 relates the two grains in the metastable G boundary configuration.

\subsection{Energetics and atom arrangements}

\begin{figure}
\caption{LDFT result for the relaxed supercell of the S interface viewed
along the [$0001$] (above) and the [$\bar{1}2\bar{1}0$] (below)
directions. The supercell contains two equivalent interfaces (dashed
lines), one in the middle, the other at the extremes arising from the
periodic boundary conditions in the [$10\bar{1}0$] direction. The atoms
are represented as in Fig.~\ref{cryst}.}
\label{sfig}
 \centerline{\psfig{file=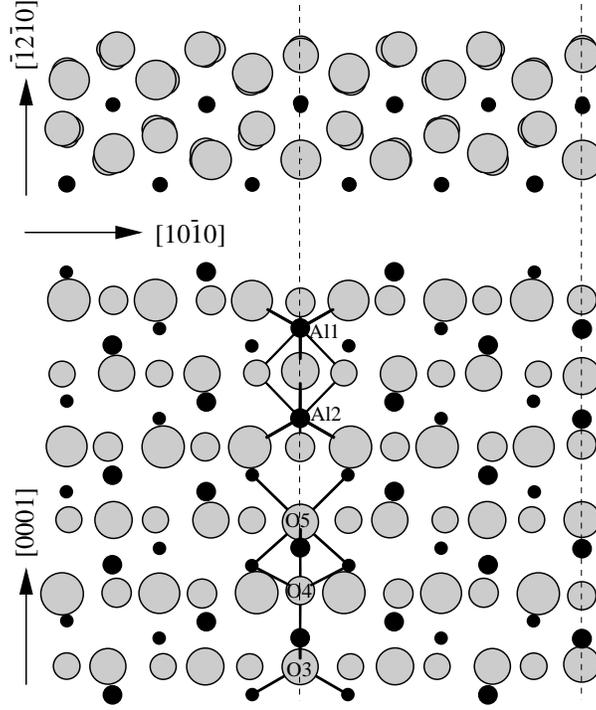,width=8.0cm,angle=0}} 
\end{figure}

\begin{figure}
\caption{LDFT result for the relaxed atomic structure of the G interface
viewed along [$0001$] (above) and the [$\bar{1}2\bar{1}0$] (below)
directions. Atoms, labels, and symbols as in Fig.~\ref{sfig}.}
\label{gfig}
 \centerline{\psfig{file=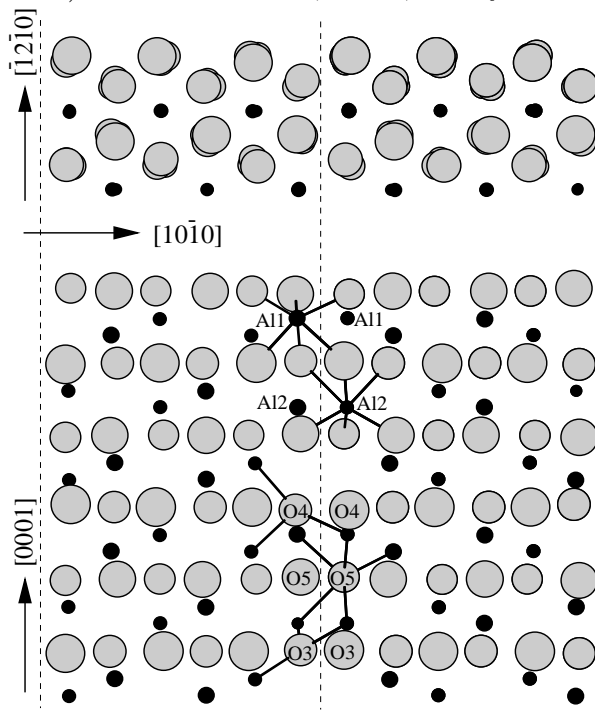,width=8.0cm,angle=0}} 
\end{figure}

 The internal degrees of freedom of the two supercells describing the S and
 G boundaries were fully relaxed with respect to the LDFT atomic
 forces. The resulting interface structures are highly ordered and are
 shown in Fig.~\ref{sfig} and Fig.~\ref{gfig}. The interface energy was
 calculated as the difference between the total energy of the relaxed
 supercell, and the total energy of an equal number of \al2o3 bulk
 formula-units, divided by the total interface area in the supercell
 (121.6 \AA$^2$).
 
 The S boundary variant is the most stable structure of the prismatic
 twins, with an interface energy of 0.30 J/m$^2$, which is the lowest
 interface energy of the alumina boundaries studied so far from
 first-principles (0.63 J/m$^2$ for the $\Sigma 7$
 rhombohedral,~\cite{Marinop00} 0.73 J/m$^2$ for the $\Sigma 3$
 basal,~\cite{Marinop01} and 1.88 for the $\Sigma 13$
 pyramidal~\cite{sigma13} twins). The metastable G variant is characterized
 by a higher interface energy, 0.49 J/m$^2$, but still very low if compared
 to the basal and rhombohedral twins. These values correspond to zero
 excess volume (no expansion along $T_3$), therefore it is expected that a
 cell-volume minimization would further reduce the calculated interface
 energies. However, the energy lowering due to the inter-granular
 separation is always very small (of the order of some mJ/m$^2$),
 particularly the one for highly ordered boundary structures. As a
 consequence, this is a case in which the relaxation with respect to the
 inter-granular separation will certainly be very small and can therefore
 safely be neglected.

 For completeness, the S and G boundary variants were also structurally
 relaxed according to the shell-model forces. The resulting structures were
 similar to the corresponding LDFT ones. The shell-model calculations
 predict the correct energy ordering between the S and G interfaces, with
 absolute values of 0.1 and 0.4 J/m$^2$, respectively. It is worth noting
 that this is not a general rule, since the shell-model predicts the
 correct qualitative energy ordering between the variants of the
 rhombohedral $\Sigma 7$ (Ref.~\onlinecite{Marinop00}) and of the $\Sigma
 13$ (Ref.~\onlinecite{sigma13}) twins, but fails in the case of the basal
 $\Sigma 3$ twin (Ref.~\onlinecite{Marinop01}).

\subsubsection{The screw-rotation twin S}

 The structure of the prismatic screw-rotation twin interface relaxed via
 LDFT is shown in Fig.~\ref{sfig}. The interface, marked by dashed lines,
 is decorated by Al and O atoms.  It is highly ordered, so much that the
 projection of the twin on the (0001) plane (upper panel) is
 indistinguishable from the one of the corundum single crystal (upper panel
 of Fig.~\ref{cryst}). If the small interfacial relaxation of the oxygen
 sublattice is neglected, the boundary does not affect the symmetry of the
 anion sublattice, that is hexagonal close-packed throughout the
 supercell. In fact, the twin interface arises mainly from a modified
 packing of the sublattice of cations, which fills two thirds of the
 octahedral interstitial sites in the anion sublattice, leaving one third
 of the sites empty (denoted as ``voids'' in the following).  The
 arrangement of the voids across the twin boundary can be clearly
 identified by the symmetric V-shaped pattern.

 The Al-O bond lengths of the interfacial atoms are included in
 Table~\ref{tabdist}, together with the Al-O distances in bulk \aal2o3
 (short b$_s$=1.84 \AA\ and long b$_l$=1.96 \AA\ bonds). The interfacial
 Al1 and Al2 atoms are six-fold coordinated, with three bond lengths
 centered around b$_s$ and three around b$_l$, hence the cation sites are
 very bulk-like.  Since the boundary affects mainly the packing of the
 cation sublattice in the boundary region, the main changes in coordination
 number and local atomic environment are shown by the interfacial oxygen
 atoms. These are distributed between an equal number of three-fold
 under-coordinated (O3), four-fold coordinated (O4), and five-fold
 over-coordinated (O5) atomic sites. The O4 anions are bulk-like
 tetrahedrally coordinated, with two short bonds centered at b$_s$ and two
 long bonds centered at b$_l$. The under-coordinated O3 oxygens have three
 short bonds in a planar geometry, while the O5 anions have one short and
 four long bonds.

 The atoms in the first atomic layer away from the interface are already
 bulk-like, with Al-O distances of 1.84 and 1.96 \AA, therefore showing
 that the perturbation in the corundum crystal symmetry due to the presence
 of the boundary is extremely confined to the interface plane, with a
 transverse extension of less than 2.5 \AA\ (compared for instance to 5-6
 \AA\ in the $\Sigma 13 (10\bar{1}4)$ twin~\cite{sigma13}).

\subsubsection{The glide-mirror twin G}

 The structure of the prismatic glide-mirror twin relaxed via LDFT is shown
 in Fig.~\ref{gfig}. The dashed line marks the position of the interface,
 that bisects two planes of atoms in this case. The boundary structure has
 many similarities with that one of the S boundary: the (0001) projection
 is bulk-like and the interface is determined by the cation sublattice
 only.  As a result, the boundary cations are six-fold bulk-like
 coordinated, while the main changes in the local environment are shown by
 the boundary anions. Similarly to the S variant, these are distributed
 between three-fold under-coordinated (O3), four-fold bulk-like coordinated
 (O4), and five-fold over-coordinated (O5) sites (Table~\ref{tabdist}). O3
 has three short bond lengths, O4 has two short and two long ones, and O5
 has one short and four long ones.  The inner atoms neighboring the
 boundary layer are in a bulk-like environment, therefore the transverse
 extension of the boundary is again very small in this case ($\leq$ 4 \AA
 ).

 The main distinctions between the two twin models can be summarized by
 these two points: (1) The stacking sequence of the inequivalent boundary
 anions along the [0001] (O3-O4-O5 in the S boundary, O5-O4-O3 in the G
 boundary), which is accessible by ELNES measurements with spatial
 single-atom-column resolution.  (2) The pattern formed by the cation-void
 sites across the boundary (V-shaped with one single void at the S
 boundary, V-shaped with a pair of voids at the G boundary), which is
 accessible by quantitative HRTEM analysis.

\subsubsection{The free prismatic surface}

 The free prismatic surface was relaxed by means of LDFT in a 60-atom
 supercell containing six layers of bulk alumina. The transverse size of
 the supercell was chosen to be big enough so that to minimize the spurious
 interaction between the free surfaces. These were separated by 9.5 \AA\ of
 vacuum and by 6.8 \AA\ of bulk corundum.

 The relaxation involves mainly the two outermost atomic layers, with minor
 atomic displacements ($\le 0.05$ \AA) in the inner third layer. The main
 features of the relaxed surface are the outward expansion (0.2 \AA ) of
 the external anions, the inward contraction (0.3 \AA ) of the external
 cations, and the large lateral displacement (0.4 \AA ) of the surface
 anions. The calculated LDFT value for the surface energy is 2.58 J/m$^2$.

\subsection{Projected densities of states and ELNES of oxygen}

 Experimentally, the boundary structure of the $\Sigma 3$ prismatic twin
 was studied by EELS, measuring the $K$-edge ELNES of oxygen.  When ELNES
 is analyzed by means of the spatial difference technique,~\cite{Bruley93}
 it gives an indirect information about the local bonding and coordination
 through a direct measure of the unoccupied electron states present in the
 neighborhood of an excited atom. These electron states can be obtained
 from the calculation of the double differential scattering cross section
 for the excitation of an atom by inelastic electron
 scattering.~\cite{Nelhiebel99} In the special case of the oxygen $1s
 \rightarrow 2p$ excitations, and in the dipole approximation for the
 electron scattering, the scattering cross section, and therefore the ELNES
 $K$-edge for an oxygen site, can directly be related to the calculated
 site- and $p$-projected density of states ($p$-PDOS, $p$ denotes the
 angular momentum $l$=1).~\cite{Brydson96,Koestlmeier99} Final-state
 (core-hole) effects are included in these calculations via the Z+1
 approximation, where the excited oxygen atom is simulated by a
 substitutional fluorine impurity atom.~\cite{Elsaesser01UM,Nufer01UM} The
 $p$-PDOS at different excited oxygen sites are calculated from the LDFT
 one-electron eigenvalues on the 2x1x1 {\bf k}-point mesh, and smoothened
 by a Gaussian of 0.8 eV to mimic instrumental and life-time broadening
 effects.

\begin{figure}
\caption{Comparison between the measured ELNES oxygen $K$-edge signal and
the calculated $p$-PDOS for the interfacial anions in the G and S
boundary variants. (The $p$-PDOS is the sum of the individual $p$-PDOS
for the three interfacial oxygen sites O3, O4, and O5).}
\label{elnes}
 \centerline{\psfig{file=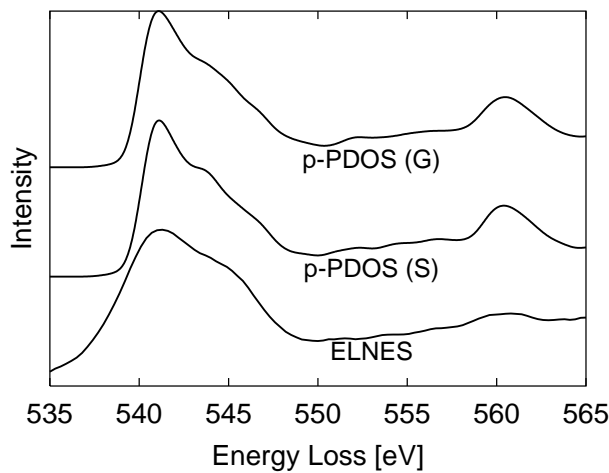,width=8.5cm,angle=0}}
\end{figure}

 To see whether the calculated lowest-energy boundary structure is the same
 as the experimentally observed one, we compare the calculated $p$-PDOS of
 the S and G interfaces with the measured ELNES $K$-edge in
 Fig.~\ref{elnes}. The experimental interface spectrum shows a pronounced
 peak at 541 eV with a shoulder at higher energies, and a minor peak around
 560 eV. The calculated $p$-PDOS curves for the S and G interfaces are
 shifted along the energy axis to match the first peaks above the edge
 onset with the peak at 541 eV in the experimental spectrum. The calculated
 $p$-PDOS are very similar to the experimental ELNES, with a good agreement
 between the relative position and relative height of the peaks. However,
 they are also very similar to each other, so that, from this comparison,
 it is not possible with sufficient confidence to discriminate one or the
 other theoretical boundary structure from the real interface.

\begin{figure}
\caption{Individual contributions D(E) of the anions O3, O4, and O5 to
the total interfacial oxygen $p$-PDOS calculated for the S (top) and G
(bottom) boundaries.}
\label{elnes2}
 \centerline{\psfig{file=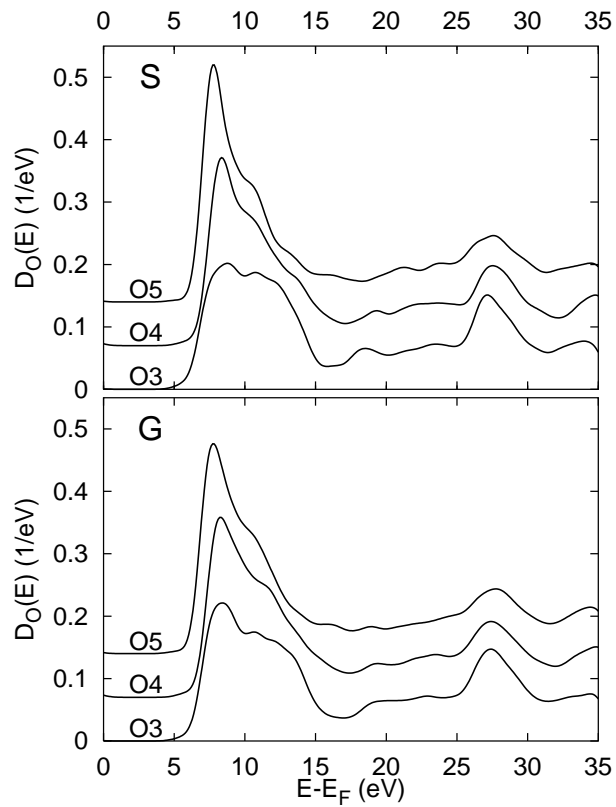,width=8.5cm,angle=0}}
\end{figure}

 The similarity between the $p$-PDOS can be understood by recalling that
 they reflect the S and G boundary atomic environments, which are
 qualitatively and quantitatively very similar
 (Table~\ref{tabdist}). Fig.~\ref{elnes2} shows the breakdown of the total
 interfacial $p$-PDOS, displayed in Fig.~\ref{elnes}, into the individual
 contributions corresponding to the O3, O4, and O5 anions. The difference
 in coordination number affects mainly the low energy region of the
 spectrum.  Over- and under-coordination drives the distinctive ``peak and
 shoulder'' shape of the bulk-like atoms, towards a single peaked, and
 double peaked shape, respectively. These distinctive signals coming from
 three-, four-, and five-fold coordinated atoms should in principle be
 measurable by EELS in a STEM with spatial single-atom-column
 resolution. In this case it would indeed be possible to discriminate
 between the two boundary structures, even though their $p$-PDOS are nearly
 indistinguishable. A line scan along the boundary layers would produce
 periodic sequences of spectra corresponding to the O3, O4, and O5
 anions. The atom-column specific signals for the S and G variants would be
 similar, but their spatial sequence along the direction $[0001]$ parallel
 to the interface would be reversed in order, thus allowing to discriminate
 between the two interfaces: ...-O3-O4-O5-... in the case of the S
 boundary, ...-O5-O4-O3-... in the case of the G boundary in bicrystals
 with equivalent bulk orientations.  However, for the time being such a
 level of spatial atom-column resolution is set as a challenge for
 experimental STEM-EELS microanalysis.~\cite{Benthem02}

\subsection{Simulated HRTEM images}

 For a bicrystal sample of \aal2o3 illuminated by a spatially extended,
 parallel electron beam along the [$\bar{1}2\bar{1}0$] zone axis in a
 high-voltage TEM with atomic-resolution capability, the main feature of
 the resulting HRTEM image is typically a strong intensity-contrast pattern
 with intensity extrema at the positions of unoccupied cation sites (cation
 void sites) in the corundum structure of \aal2o3.  The detailed
 atom-column resolved projection of the bicrystal structure is not directly
 accessible from the measured HRTEM micrograph.  It can be deduced by
 comparing it via quantitative matching with simulated HRTEM images from
 atomistic-model candidate
 structures.~\cite{Moebus94,Moebus98,Stadelmann87} For optimum agreement,
 all atom-column positions as well as imaging parameters (e.g., defocus,
 specimen thickness, lens aberrations and transfer function of the
 recording medium) need to be determined variationally.

\begin{figure}
\caption{Simulated HRTEM images of the S and G boundaries. Dashed lines
mark the rhombohedral planes.}
\label{hrtem}
 \centerline{\psfig{file=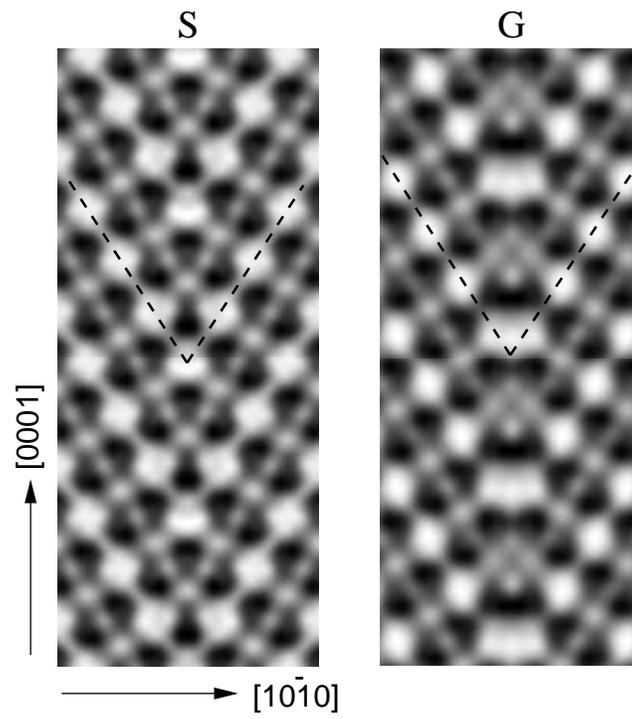,width=8.5cm,angle=0}}
\end{figure}

 As discussed above, we have extracted two metastable structures which
 should in principle result in different HRTEM images in the
 ($\bar{1}2\bar{1}0$) projection. The simulated HRTEM images of the S and G
 supercells corresponding to a TEM sample thickness of 6 nm and to a
 defocus value of $-$20 nm, which are typical parameters for HRTEM of
 \aal2o3 in an instrument like the JEM ARM-1250 at the MPI f\"{u}r
 Metallforschung Stuttgart,~\cite{Phillipp94,GemmingUnp,RichterUnp} are
 displayed in Fig.~\ref{hrtem}. The two images show characteristic features
 that allow a discrimination between the two interfaces.  In the S variant,
 the white spots located at the boundary plane are visually equivalent to
 the ones inside the bulk grains. In the G variant, the interfacial white
 spots have a distinctively oversized spot, arising from two neighboring
 spots merging together. The difference in the micrographs does not arise
 from the interfacial translation states, since both the S and G variants
 produce a symmetric V-shaped pattern of octahedral void sites, but from
 the local boundary structure. In particular, the rhombohedral planes of
 voids in different grains intersect the interface plane precisely at one
 cation-void site in the S variant (Fig.~\ref{sfig}), and on an oxygen
 basal plane in the G variant (Fig.~\ref{gfig}). This is schematically
 shown by the dashed lines in Fig.~\ref{hrtem}, where the cation-void sites
 are associated to the white spots.  These images suggest that an
 high-resolution micrograph would discriminate between the two boundary
 variants. Experimental HRTEM images~\cite{NuferTh,GemmingUnp} show a
 contrast pattern compatible with the simulated one for the S interface,
 apparently confirming the LDFT prediction based on the lowest interface
 energy. Hence, for such experimental images a quantitative HRTEM
 analysis~\cite{Moebus94,Moebus98,Stadelmann87} starting from our
 calculated two interface models appears promising
 (cf. Ref.~\onlinecite{Nufer01}).

\section{Conclusions}

 In this work, the energetic stabilities, atomistic arrangements and
 electronic states at the prismatic $\Sigma$3 ($10\bar{1}0$) twin interface
 in \aal2o3 were analyzed theoretically by atomistic shell-model and
 ab-initio LDFT calculations, and experimentally by spatially resolved
 STEM-EELS measurements of the oxygen $K$-ionization-edge ELNES.

 Two metastable microscopic variants for the prismatic twin with low
 interfacial energies resulted from the theoretical analysis. They exhibit
 subtle differences in the local arrangements of atoms at their interfaces,
 which should be distinguishable via TEM bicrystal experiments.

 Experimentally, it was demonstrated that a state-of-the-art STEM
 instrument, with spatial resolution of 1$-$2~nm and energetical resolution
 of about 0.8 eV, is not yet capable to resolve the subtle differences in
 the interfacial ELNES of this very bulk-like twin interface of
 \aal2o3. However, with the next generation of analytical TEM instruments,
 which are designed to achieve resolutions of 0.1$-$0.2~nm in space and
 about 0.2 eV in energy and currently approaching the stage of operation,
 the two theoretical twin variants will become distinguishable.

 Hence, it is proposed that the present theoretical ab-initio results for
 the prismatic twin in \aal2o3 provide a very promising benchmark case for
 the coming analytic TEM instruments: (a) to see whether single-atom-column
 resolution is really achieved by detecting significant differences in the
 oxygen $K$-edge ELNES for the three distinct interfacial oxygen sites; (b)
 to check whether the ab-initio LDFT prediction of the S variant with
 lowest interface energy is also the experimentally observed case. For the
 ELNES experiment, the spatial sequence of the three oxygen sites is
 decisive. Alternatively, for a HRTEM experiment combined with a
 quantitative image-matching analysis, the contrast patterns originating
 from arrangements of cation-void sites across the interfaces discriminate
 the two theoretical variants.

\acknowledgments

 This work was supported by the Deutsche Forschungsgemeinschaft (Project
 No. El 155/4-1).  The authors thank Dr. Thomas Gemming for communicating
 his experimental HRTEM images of a prismatic twin interface prior to
 publication, Dr. Gunther Richter for his help with the simulation of HRTEM
 images from the two theoretical twin-interface models, and Prof. Manfred
 R\"{u}hle for his continuous interest and support.

\begin{table}
\caption{Equilibrium Al-O bond lengths (in \AA) in the interfacial regions
of the S and G supercells. For comparison, the Al-O distances in bulk
\aal2o3 are also included. Atoms are numbered as in Fig.s 4 and 5}
\label{tabdist}
\begin{center}
\begin{tabular}{c|cccccccc}
        & \multicolumn{8}{c}{\it Bulk} \\
Al-O    & 1.84 & 1.84 & 1.84 & \hspace{0.5cm} & 1.96 & 1.96 & 1.96 & 
                                                        \\ \hline
        & \multicolumn{8}{c}{\it Screw twin} \\
Al1-O   & 1.79 & 1.86 & 1.86 & & 1.95 & 1.95 & 2.04 & \\
Al2-O   & 1.83 & 1.84 & 1.84 & & 1.90 & 1.98 & 1.98 & \\
O3-Al   & 1.76 & 1.76 & 1.79 & &      &      & & \\
O4-Al   &      & 1.83 & 1.83 & & 1.90 & 2.04 & & \\
O5-Al   &      &      & 1.83 & & 2.03 & 2.04 & 2.09 & 2.10 \\ \hline
        & \multicolumn{8}{c}{\it Glide twin} \\
Al1-O   & 1.80 & 1.81 & 1.84 & & 1.92 & 1.94 & 2.08 & \\
Al2-O   & 1.76 & 1.76 & 1.83 & & 1.91 & 2.04 & 2.12 & \\
O3-Al   & 1.76 & 1.76 & 1.76 & &      &      & & \\
O4-Al   &      & 1.81 & 1.84 & & 1.93 & 1.96 & & \\
O5-Al   &      &      & 1.88 & & 1.94 & 2.05 & 2.08 & 2.12 \\ 
\end{tabular}
\end{center}
\end{table}


\newpage

\newpage

\newpage

\newpage

\newpage

\end{document}